\documentclass[12pt]{article}
\usepackage[utf8]{inputenc}
\usepackage{mathtools,amsmath,amssymb}
\usepackage{graphicx}
\usepackage{lmodern}
\usepackage[T1]{fontenc}
\usepackage{microtype}
\usepackage[pagebackref=false]{hyperref}
\numberwithin{equation}{section}
\usepackage{dsfont} 
\usepackage{amsthm}

\usepackage[textsize=tiny]{todonotes}

\usepackage{xspace}
\usepackage{bibentry}
\usepackage{easybmat}

\usepackage{graphicx}

\usepackage{subcaption} 

\usepackage[numbers,sort&compress]{natbib}
\usepackage{hypernat}
 
\usepackage{tikz}
\usetikzlibrary{decorations.pathreplacing} 
 
\usepackage[margin=2.5cm]{geometry}
    
\newcommand{\NN}{\mathbf{N}}

\newcommand{\rang}{\rangle\hspace{-0.1cm}\rangle}
\newcommand{\lang}{\langle\hspace{-0.1cm}\langle}

\newmuskip\pFqmuskip

\newcommand*\pFq[6][8]{%
  \begingroup 
  \pFqmuskip=#1mu\relax
  \mathchardef\normalcomma=\mathcode`,
  \mathcode`\,=\string"8000
  \begingroup\lccode`\~=`\,
  \lowercase{\endgroup\let~}\pFqcomma
  {}_{#2}F_{#3}{\left[\genfrac..{0pt}{}{#4}{#5};#6\right]}%
  \endgroup
}
\newcommand{\pFqcomma}{{\normalcomma}\mskip\pFqmuskip}

\newcommand{\oa}{\mathbf{a}}

\newcommand{\oad}{\mathbf{\bar{a}}}

\begin{document}

\begingroup
\begin{center}

\vspace{6em}
 \begingroup\LARGE
 \bf Steady state representations for the harmonic process
\par\endgroup
 \vspace{3.5em}
 \begingroup\large \bf
  Rouven Frassek
 \par\endgroup
\begin{center} 
 \end{center}

\vspace{1em}

\begingroup\sffamily
\small
Dipartimento di Scienze Fisiche, Informatiche e Matematiche, Università di Modena e Reggio
Emilia, Via G. Campi 213/B, 41125 Modena and INFN, Sezione di Bologna, Via Irnerio 46,
40126 Bologna, Italy
\par\endgroup
\vspace{3em}

\end{center}
 \begin{abstract}
 \noindent
In this note we discuss how the matrix product solution for the steady state of the   harmonic process is obtained from the solutions already known in the literature, i.e. the   closed-form expression derived in \cite{Frassek:2021yxb} and the nested integral form obtained in \cite{2024JSP...191...10C,Carinci:2023nfn}. Our results  clarify the relation between the three representations of the steady state and provide the matrix product solution that has not been available for this model before.
\end{abstract}

\vspace{4em}







\section{Introduction}

The matrix product ansatz \cite{Derrida:1992vu} remains one of the most elegant methods to describe ground states of Hamiltonians. Its origins are deeply rooted in the theory of quantum integrable systems and the quantum inverse scattering method. By now it is rather well understood  \cite{sasamo,Crampee,2015JPhA...48L4001C} that the construction is closely related to the Zamolodchikov and Goshal-Zamolodchikov algebra of 2d integrable quantum field theories \cite{Zamolodchikov:1978xm,Ghoshal:1993tm}.

In stochastic particle processes like the SSEP and ASEP, the matrix product ansatz has been proven to be powerful to obtain steady states \cite{Derrida:1992vu}, see also \cite{schutz1993phase} for an alternative method, but can also be applied to the open Heisenberg spin chain \cite{Prosen:2011wmh} and other integrable spin chains, see e.g.~\cite{2017JSP...166.1129V}.  In the case of the ASEP and SSEP, each site of the chain can either be empty or occupied by one single particle. These two states are respectively represented by an operator in the matrix product ansatz satisfying a certain so called DEHP algebra that is equivalent to the Zamolodchikov and Goshal-Zamolodchikov algebra. In general the representations of this algebra are infinite-dimensional and finding representations of the algebra can be an involved problem.

The situation is more difficult if the stochastic system allows to accommodate an unbounded number of particles at each site. Such integrable particle processes where studied by Sasamoto and Wadati in \cite{Sasamoto}, Povolotsky in \cite{Povolotsky} and Barraquand and Corwin in \cite{barraquand}. Later on the integrable processes were extended with boundaries in \cite{Frassek:2022fjs}. In the following we will focus on the symmetric (rational) limit of this process that was defined in \cite{Frassek:2019vjt} and is referred to as the harmonic process. It arises from the nearest-neighbor Hamiltonian of the open Heisenberg XXX spin chain with non-compact spin representations in the quantum space and a certain choice of boundary conditions.

Given that the configuration space of the harmonic model is unbounded,  we expect that the matrix product algebra that is relevant for this case has infinitely many generators. Due to this difficulty, the authors in \cite{Frassek:2021yxb} developed a method to obtain the steady state from the quantum inverse scattering method \cite{Faddeev:1996iy} using the charges of the spin chain. The same strategy works for the finite-dimensional SSEP \cite{Frassek:2019imp,Frassek:2020omo}, see also \cite{49baa546-5bb8-3d6e-8adc-8b5403de5cf6}. Later on, a less explicit but rather concise expression in terms of nested integrals was obtained for the steady state of the harmonic process in \cite{2024JSP...191...10C,Carinci:2023nfn} which can be interpreted as a mixed measure. The authors of \cite{Frassek:2023gji} showed that such expressions can also generalised for the q-deformed version and also noticed the similarity with the matrix product ansatz.

In this note we like to fill the gap and provide a matrix product ansatz for the harmonic process. As we will show, a matrix product solution can be obtained from the closed-form expression of the steady state as well as  from its integral representation that were known already. The resulting matrix product representations of the stationary state $|\mu\rangle$ that satisfies
\begin{equation}\label{eq:steadc}
 H|\mu\rangle =0\,,
\end{equation}
are then of the form
\begin{equation}
 |\mu\rangle =\sum_{m_1,\ldots,m_N=0}^\infty \mu(m_1,\ldots,m_N)|m_1,\ldots,m_N\rangle\,,
\end{equation}
with the components of the canonical basis $|m_1,\ldots,m_N\rangle$, see \eqref{eq:sl2ac}, given in matrix product form
\begin{equation}\label{eq:prodf}
 \mu(m_1,\ldots,m_N) = Z_N^{-1}\lang V|X(m_1)\cdots X(m_N)|W\rang\,.
\end{equation}
Here $N$ denotes the length of the chain, the variables  $m_i\in \mathbb{N}$ count the number of particles per site, $Z_N$ is a normalisation and $X(m)$ are the generators of the matrix product algebra in the bulk and $|W\rang$ and $\lang V|$ are boundary states that act in the auxiliary space of the matrix product state.

The article is organised as follows. In Section~\ref{sec:harm} we introduce the harmonic process with spin label $2s\in\mathbb{N}$ and review the two representations known for the steady state. In Section~\ref{sec:mps} we discuss the matrix product representation of the steady state. More precisely we discuss how it can be obtained from the known representations and provide an independent proof that the matrix product algebra is satisfied. Finally we end with a conclusion in Section~\ref{sec:conc}.

\section{The harmonic process and its solution}\label{sec:harm}
In this section we define the harmonic process and present the two representations of the steady state that are known so far.

As mentioned above, the stochastic Hamiltonian of the  harmonic process
is identical to  the nearest-neighbor Hamiltonian of the Heisenberg spin chain  with non-compact representations of the $sl(2)$ Lie algebra
\begin{equation}
\label{eq:sl2 discrete}
 [S_0,S_\pm]=\pm S_\pm\,,\qquad [S_+,S_-]=-2S_0\,.
\end{equation}
More precisely we are interested in the representations
that act on the Hilbert space as follows
\begin{equation}\label{eq:sl2ac}
  S_+|m\rangle = (m+2s)|m+1\rangle\,,\qquad  S_-|m\rangle  =  m|m-1\rangle\,,\qquad S_0|m\rangle  =  (m+s)|m\rangle\,,
\end{equation}
where in the following we assume that  $2s\in \mathbb{N}$ and $m\in \mathbb{N}$.\footnote{{This representation can be made unitary when considering the similarity transformation $ K_\pm=DS_{\pm}D^{-1}$ and $K_0=S_0$ where the entries of the diagonal matrix $D$ are $ D_m=\sqrt{\frac{ \Gamma (m+1) \Gamma (2 s)}{\Gamma (m+2 s)}}$. }}

The configuration space of the harmonic process is defined on the $N$-fold tensor product of such $sl(2)$ modules and it is convenient  to define $N$ copies of generators $S_a^{[i]}$ with $a=\pm,0$ and $i=1,\ldots,N$ that act on the $i$th term in the tensor product. The total spin generators then take the form
\begin{equation}\label{stot}
 S_a^{tot}=\sum_{i=1}^N S_a^{[i]}\,.
\end{equation}

\subsection{The stochastic Hamiltonian}
The dynamics of the harmonic process is governed by the stochastic Hamiltonian. It is of nearest-neighbor type with boundary terms that act as reservoirs on the first and last site indicated by $\mathcal{B}_{L,R}$ respectively. We have
\begin{equation}\label{eq:fullham-sss}
H=\mathcal{B}_{L}+\sum_{i=1}^{N-1}\mathcal{H}_{i,i+1}+\mathcal{B}_R\,.
\end{equation} 
The bulk part is formed out of the Hamiltonian density that acts on the tensor product of two sites as
\begin{equation}
\label{eq:hacts-ss}
\begin{split}
 \mathcal{H}|m\rangle\otimes|m'\rangle=\left(h_s(m)+h_s(m')\right)|m\rangle\otimes|m'\rangle&-\sum_{k=1}^{m}
 \varphi_s({m},{k})
 |m-k\rangle\otimes|m'+k\rangle
 \\&-\sum_{k=1}^{m'}
  \varphi_s(m',{k})
 |m+k\rangle\otimes|m'-k\rangle\,,
 \end{split}
\end{equation}
with
\begin{equation}
 h_s(m)=\sum_{k=1}^m\frac{1}{2s+k-1}\,.
\end{equation}
 The action of the Hamiltonian density was obtained in \cite{Martins:2009dt} while the interpretation as jump rates of a continuous-time Markov process was given in \cite{Frassek:2019vjt}. The rates in \eqref{eq:hacts-ss} are given by
\begin{equation}
 \label{eq:varphi}
\varphi_s(m,k) =  \frac{1}{k}\frac{\Gamma (m+1) \Gamma (m-k+2 s)}{ \Gamma (m-k+1) \Gamma (m+2 s)}\,.
\end{equation}
In the Hamiltonian \eqref{eq:fullham-sss}, the term $\mathcal{H}_{i,i+1}$ denotes the Hamiltonian density acting non-trivially on site $i$ and $i+1$.

At the boundaries of the chain, i.e. at site $1$ and site $N$, particles are extracted and inserted. The boundary terms read
\begin{equation}\label{eq:harmbndss}
  \mathcal{B}_{L,R}|m\rangle=\left(h_s(m)+\sum_{k=1}^\infty\frac{\beta_{L,R}^k}{k}\right)|m\rangle -\sum_{k=1}^{m}
   \varphi_s({m},{k})
|m-k\rangle -\sum_{k=1}^\infty \frac{\beta_{L,R}^k}{k}|m+k\rangle\,,
\end{equation}
where the insertion rates are determined through two parameters $0<\beta_{L,R}<1$.

\subsection{Closed-form expression of the  stationary state}\label{sec:clos}

The stationary state of the Hamiltonian \eqref{eq:fullham-sss}  satisfying \eqref{eq:steadc} has been found in \cite{Frassek:2021yxb}. It can compactly be written in terms of two global rotations in terms of the total spin generators \eqref{stot} that naturally arise from the symmetries of the process and the algebraic structure of the boundary operators, see also the discussion at the beginning of Section~\ref{sec:mps}. Using the quantum inverse scattering method one  finds that
\begin{equation}\label{stat}
 |\mu\rangle =\exp\left[-S_-^{tot}\right]\exp\left[{\rho_R} S_+^{tot}\right]|\nu\rangle\,,
\end{equation}
where the non-trivial structure of the steady state is contained only in
\begin{equation}
|\nu\rangle=\sum_{m_1,\ldots,m_N=0}^\infty \nu(m_1,\ldots,m_N)|m_1,\ldots,m_N\rangle\,.
\end{equation}
Here we introduced the variable
\begin{equation}
 \rho_{L,R}=\frac{\beta_{L,R}}{1-\beta_{L,R}}\,.
\end{equation}
The components of the non-trivial part of \eqref{stat} can be written as  telescopic products. They read
\begin{equation}
\label{eq:mupp}
\nu(\vec m) =
(\rho_{L}-\rho_{R})^{|\vec m|} \frac{\Gamma(2s (N+1))}{\Gamma(|\vec m|+2s(N+1))} \prod_{i=1}^N\kappa(m_i)
\frac{\Gamma\left(2s(N+1-i)+\sum_{k=i}^{N}m_k\right)}{\Gamma\left(2s(N+1-i)+\sum_{k=i+1}^{N}m_k\right)}\,,
\end{equation}
with $\vec m=(m_1,\ldots,m_N)$ and
\begin{equation}\label{eq:kappa}
 \kappa(m)=\frac{\Gamma(2s+m)}{\Gamma(2s)\Gamma(1+m)}\,.
\end{equation}
This expression has been obtained by acting with a non-local charge of the spin chain that has been derived within the quantum inverse scattering method on the reference state $|0,\ldots,0\rangle$.

\subsection{Integral representation of the steady state}

The closed-form expression of the steady state above was written as nested integral in \cite{2024JSP...191...10C,Carinci:2023nfn} which can be interpreted as a mixed measure.
For general spin values the expression obtained is given by
\begin{equation}\label{mixtresult-explicit1}
 \begin{split}
\mu(\vec m)
=
\frac{1}{(\rho_r-\rho_l)^{2s(N+1)-1}}&\frac{\Gamma(2s(N+1))}{\Gamma(2s)^{N+1}}
 \int_{\rho_L}^{\rho_R}d\theta_1 \int_{\theta_1}^{\rho_R}d\theta_2\cdots  \int_{\theta_{N-1}}^{\rho_R}d\theta_N    \\
& \qquad  \times\Big[\prod_{i=1}^{N+1} (\theta_{i}-\theta_{i-1})^{2s-1}\Big]
\Big[\prod_{i=1}^{N} \kappa(m_i)\left(\frac{ \theta_i}{1+ \theta_i}\right)^{m_i} \left(\frac1{1+ \theta_i}\right)^{2s}\Big]\,.
\end{split}
\end{equation}
In particular, as shown in \cite{Carinci:2023nfn},  when applying the transformation in \eqref{stat} one finds that
\begin{equation}\label{mixnu}
 \begin{split}
  \nu(\vec m)
=
\frac{(-1)^{(2s+1)N}}{u_0^{2s(N+1)-1}}\frac{\Gamma(2s(N+1))}{\Gamma(2s)^{N+1}}
 \int_{0}^{u_0}du_1 & \int_{0}^{u_1}du_2\cdots  \int_{0}^{u_{N-1}}du_N     \\& \times u_{N}^{2s-1} \Big[\prod_{i=1}^{N}\kappa( m_i)u_i^{\eta_i} (u_{i}-u_{i-1})^{2s-1}\Big]\,,
 \end{split}
\end{equation}
where we introduced the variables $u_i=\theta_i-\rho_R$ and $u_0=\rho_L-\rho_R$.

Then to show that \eqref{mixnu} coincides with the closed-form expression \eqref{eq:mupp}, the main ingredient is the integral representation of the beta function. After the transformation $t=y/x$ it be brought to the form
\begin{equation}
 B(a+1,b+1)
 =\int_0^1dt\, t^a(1-t)^b=
 (-1)^b x^{-a-b-1}\int_{0}^x dy\, y^a (y-x)^b\,.
\end{equation}
This relation allows to perform the nested integrals. Evaluating the first one yields
\begin{equation}\label{mixtresult-explicit4}
 \begin{split}
\nu(\vec m)
= \left[\prod_{i=1}^N\kappa( m_i)\right]
\frac{\Gamma(2s(N+1))}{\Gamma(2s)^{N+1}}\frac{(-1)^{2s-1}}{u_0^N}&B(2s+\eta_N,2s)
 \int_{0}^{u_0}du_1 \int_{0}^{u_1}du_2\cdots  \int_{0}^{u_{N-2}}du_{N-1}    \\
&    \qquad \times (u_{N-1})^{4s-1+m_N} \left[\prod_{i=1}^{N-1}u_i^{m_i} (u_{i}-u_{i-1})^{2s-1}\right]\,,
 \end{split}
\end{equation}
while after $N$ steps of iteration we obtain
\begin{equation}
\begin{split}
\nu(\vec m)
 &=(\rho_{L}-\rho_{R})^{|\vec m|}
\frac{\Gamma(2s(N+1))}{\Gamma(2s)^{N+1}}\cdot\prod_{i=1}^N \kappa(m_i)  B(2s(N-i+1)+\sum_{k=i}^Nm_{k},2s)\,.
 \end{split}
\end{equation}
Finally using the representation of the beta function in terms of gamma functions
\begin{equation}
 B(a,b)=\frac{\Gamma(a)\Gamma(b)}{\Gamma(a+b)}\,,
\end{equation}
we arrive at
\begin{equation}\label{eq:nu2}
\begin{split}
\nu(\vec m)&=
(\rho_{L}-\rho_{R})^{|\vec m|}\frac{\Gamma(2s (N+1))}{\Gamma(2s)}
 \prod_{i=1}^N\kappa(m_i) \frac{\Gamma\left(2s(N+1-i)+\sum_{k=i}^{N}m_k\right)}{\Gamma\left(2s(N+2-i)+\sum_{k=i}^{N}m_k\right)}\,.
 \end{split}
\end{equation}
This can be shown to coincide with \eqref{eq:mupp} when shifting the index of the product in the denominator, i.e.~$i\to i-1$.

\section{Matrix product solution}\label{sec:mps}
In this section we derive a matrix product solution for the steady state of the harmonic process.

As discussed in Section~\ref{sec:clos}, the non-trivial structure of the steady state is contained in the vector $|\nu\rangle$. This is the motivation to first study  the  Hamiltonian
\begin{equation}\label{eq:htilde}
 \tilde H= \exp\left[-{\rho_R}  S_+^{tot}\right]\exp\left[S_-^{tot}\right] H\exp\left[-S_-^{tot}\right]\exp\left[{\rho_R} S_+^{tot}\right]\,,
\end{equation}
that is isospectral to the stochastic Hamiltonian in \eqref{eq:fullham-sss}. The matrix product solution of the stochastic Hamiltonian is then obtained from the one of $\tilde H$ in Section~\ref{sec:sim}. As follows from \eqref{stat}, the  eigenstate of the Hamiltonian \eqref{eq:htilde} with vanishing eigenvalue will be $|\nu\rangle$ as given in \eqref{eq:mupp} or equivalently \eqref{mixnu}, i.e. we have that
\begin{equation}\label{eq:evnu}
 \tilde H |\nu\rangle=0\,.
\end{equation}
Here the  transformed Hamiltonian \eqref{eq:htilde} is of the form
\begin{equation}\label{eq:fullham-b}
\tilde H=\tilde{\mathcal{B}}_{L}+\sum_{i=1}^{N-1}{\mathcal{H}}_{i,i+1}+\tilde{\mathcal{B}}_R\,.
\end{equation}
We remark that the bulk part is not affected by the transformation, while the boundary terms become much simpler than in the original process, cf.~\eqref{eq:harmbndss}, and read
\begin{equation}\label{eq:harmbndsss}
  \tilde{\mathcal{B}}_{L}|m\rangle=h_s(m)|m\rangle  -\sum_{k=1}^\infty \frac{(\rho_{L}-\rho_R)^k}{k}|m+k\rangle\,,\qquad \tilde{\mathcal{B}}_{R}|m\rangle=h_s(m)|m\rangle\,.
\end{equation}
For more details we refer the reader to \cite{Frassek:2021yxb}.

The matrix product form of $|\nu\rangle$ then is then inherited from the matrix product form of $|\mu\rangle$ as alluded to in \eqref{eq:prodf}. It can be written as
\begin{equation}\label{eq:mps}
 |\nu\rangle =Z_N^{-1}\lang V|Y\otimes Y\otimes \ldots\otimes Y|W\rang\,,
\end{equation}
where
\begin{equation}
 Y=\exp\left[-{\rho_R} S_+\right]\exp\left[S_-\right] X\,,
\end{equation}
and $X=\sum_{m=0}^\infty X(m)|m\rangle$ are the generators of the matrix product algebra, cf.~\eqref{eq:prodf}.
The operators $Y$ then satisfy a matrix product algebra that is governed by the equations
\begin{equation}\label{hyy}
 \mathcal{H}(Y\otimes Y)=Y\otimes \bar Y-\bar Y\otimes Y\,,
\end{equation}
and
\begin{equation}\label{bndmpa}
 \tilde{\mathcal{B}}_L \lang V|Y=\lang V|\bar Y\,,\qquad \tilde{\mathcal{B}}_R Y|W\rang =-\bar Y|W\rang\,.\
\end{equation}
Here $\bar Y$ is an auxiliary operator that is not contained in the matrix product form of the state \eqref{eq:mps}. Given these conditions, it immediately follows that \eqref{eq:evnu} is satisfied.

\subsection{From closed-form expression to MPA}
We now show how the matrix product state \eqref{eq:mps} can be constructed from the closed-form expression \eqref{eq:mupp}. For this purpose it is convenient to introduce a pair of oscillators
\begin{equation}\label{eq:osc}
 [\oa,\oad]=1\,,
\end{equation}
and a corresponding Fock space $|m\rang$. The action of the creation and annihilation operators is
\begin{equation}
 \oad |m\rang=|m+1\rang\,,\qquad \oa|m\rang =m|m-1\rang\,,
\end{equation}
while on the dual basis with $\lang m|n\rang=\delta_{m,n}$ we have
\begin{equation}
 \lang m|\oad = \lang m-1|\,,\qquad   \lang m|\oa =(m+1) \lang m+1|\,.
\end{equation}
Let us focus on the form \eqref{eq:nu2} of $|\nu\rangle$.
The telescopic form of the product yields to the following proposal:
\begin{equation}\label{eq:YY}
\begin{split}
  Y(m)
 &=\kappa(m)\oad^{2s}\frac{\Gamma\left(\NN +1 \right)}{\Gamma\left(\NN +2s+1\right)}\oad^{m}\,,
 \end{split}
\end{equation}
with $ \NN=\oad\oa$ and $\kappa(m)$ given in \eqref{eq:kappa}.
For the boundary states we fix
\begin{equation}\label{eq:bndvec}
     \lang V|=\sum_{k=0}^\infty (\rho_{L}-\rho_{R})^{k} \lang k|\,,\qquad |W\rang= \oad^{2s-1}|0\rang\,.
\end{equation}
Equivalently using matrix elements we can write the bulk operator as
\begin{equation}\label{eq:matel}
 \lang a|Y(m)|b\rang =\kappa(m)\delta_{a,2s+m+b}\frac{\Gamma(m+b+1)}{\Gamma(2s+m+b+1)}\,,
\end{equation}
and the boundary states as
\begin{equation}\label{eq:VW}
 \lang V|a\rang = (\rho_L-\rho_R)^a\,,\qquad \lang a|2s-1\rang =\delta_{2s-1,a}\,.
\end{equation}

We now verify our proposal in the following. First we note that the action of the matrix product operators simply yields
    \begin{equation}
        Y(m)|k\rang =\kappa(m)\frac{\Gamma\left(k +m+1 \right)}{\Gamma\left(k +2s+m+1\right)} |2s+m+k\rang\,,
    \end{equation}
    such that for the product of $N$ operators one obtains
\begin{equation}
\begin{split}
        Y(m_1)\cdots Y(m_N)|W\rang
        &=\left(\prod_{i=1}^N\kappa(m_i)\frac{\Gamma\left(2s(N-i+1) +\sum_{j=i}^N m_j\right)}{\Gamma\left(2s(N-i+2) +\sum_{j=i}^N m_j\right)}\right) |2s(N+1)+|\vec m|-1\rang\,.
\end{split}
    \end{equation}
    Finally, applying the left bra-vector we thus find that
\begin{equation}
\begin{split}
        &\lang V|Y(m_1)\cdots Y(m_N)|W\rang   = (\rho_L-\rho_R)^{2s(N+1)+|\vec m|-1}\left(\prod_{i=1}^N\kappa(m_i)\frac{\Gamma\left(2s(N-i+1) +\sum_{j=i}^N m_j\right)}{\Gamma\left(2s(N-i+2) +\sum_{j=i}^N m_j\right)}\right)\,.
\end{split}
    \end{equation}
This result also fixes our normalisation constant in \eqref{eq:mps}. We find
\begin{equation}
 Z_N^{-1}=\frac{1}{
(\rho_{L}-\rho_{R})^{2s(N+1)-1}}\frac{\Gamma(2s (N+1))}{\Gamma(2s)}\,.
\end{equation}
{For the case of equal densities $\rho_L=\rho_R$ we observe that the only non-vanishing component of $\nu(\vec m)$ comes from the term with $m_1=\ldots=m_N=0$.
}

\subsection{From Integral representation to MPA}

The matrix product state \eqref{eq:mps} can also be extracted from the integral representation of the steady state. For this purpose it is convenient to introduce the integral operators
\begin{equation}
 A_m[h(t)]=\frac{\kappa(m)}{\Gamma(2s)}\int_{0}^{t}dr\,
    h(r) r^m (t-r)^{2s-1}\,.
\end{equation}
Using this definition we see that the integral expression of $|\nu\rangle$ in \eqref{mixnu} can be written  as
\begin{equation}
  \nu(\vec m)=
\frac{\Gamma(2s(N+1))}{\Gamma(2s)}\frac{1}{(\rho_l-\rho_r)^{2s(N+1)-1}}(V(\rho_L-\rho_R)|\circledast A_{m_1}\circledast\ldots \circledast A_{m_N}\circledast |W(t))\,.
\end{equation} 
Here we used the notation $A_m\circledast h(t)=A_m[h(t)]$ and introduced the 'boundary vectors' $(V(\rho_L-\rho_R)|$ and $|W(t))$ that are explained in the following. The right vector is just a function
\begin{equation}
 |W(t))= t^{2s-1}\,,
\end{equation}
while the left vector is represented by an integral operator that contains the Dirac delta function
\begin{equation}
    ( V(\rho_L-\rho_R)|\circledast h(t) =\int_{-\infty}^\infty \delta(r-(\rho_L-\rho_R))h(r)dr= h(\rho_L-\rho_R)\,.
\end{equation}
The matrix form of the integral operators introduced here can then be obtained when acting on monomials and picking the pole at zero
\begin{equation}
 \left(A_m\right)_{ab}=\oint dt \frac{A_m[t^b]}{t^{a+1}}\,.
\end{equation}
More precisely, for $2s\in \mathbb{N}$ we have
\begin{equation}
  A_m[t^k]=\frac{\kappa(m)}{\Gamma(2s)} \int_{0}^{t}dr\,
    r^{m+k} (t-r)^{2s-1}=\frac{\kappa(m)}{\Gamma(2s)}t^{2s+m+k}B(m+k+1,2s)\,,
\end{equation} 
so we obtain the matrix form
\begin{equation}
\left(A_m\right)_{ab}=\frac{\kappa(m)}{\Gamma(2s)}\delta_{a,2s+m+b}B(m+b+1,2s)=\kappa(m)\delta_{a,2s+m+b}\frac{\Gamma(m+b+1)}{\Gamma(2s+m+b+1)}\,,
\end{equation}
acting on the space of monomials.
This coincides with the component form \eqref{eq:matel} obtained in the previous subsection obtained from the oscillators. Further we note that for the boundary states we have
\begin{equation}
 |W(t))_a=\oint dt\frac{ |W(t))}{t^{a+1}}=\delta_{a,2s-1}\,,
\end{equation}
and
\begin{equation}
 (V(\rho_L-\rho_R)|_a=(V(\rho_L-\rho_R)|\circledast t^a=(\rho_L-\rho_R)^a\,.
\end{equation}
Also this is in agreement with the components that arise from the oscillator representation, c.f.~\eqref{eq:VW}.
\subsection{Matrix product algebra}
In this section {we present a non-trivial representation of the matrix product algebra that arises from \eqref{hyy} and \eqref{bndmpa} using a pair of oscillators, cf.~\eqref{eq:osc}.} While the operator $Y$ is represented as in  \eqref{eq:YY}, the auxiliary operator $\bar Y$, that does not appear in the matrix product state $|\nu\rangle$, can be written as
\begin{equation}\label{ybar}
 \bar Y  =\sum_{m=1}^\infty\left(  Y(m) h_s(m) -\sum_{p=0}^{m-1} \frac{\oad^{m-p}}{m-p}Y(p)\right)|m\rangle\,.
\end{equation}
{In the following we show that this choice of $Y$ and $\bar Y$ ensures the commutation relations \eqref{hyy} and is also compatible with \eqref{bndmpa} when using the boundary vectors defined in \eqref{eq:bndvec}.
}

\paragraph{Bulk} We begin with {the verification of} the bulk relation \eqref{hyy}. For the computation it is convenient to split the Hamiltonian density into a left and right moving part $\mathcal{H}=\mathcal{H}_{right}+\mathcal{H}_{left}$ that act as
\begin{equation}
\begin{split}
 \mathcal{H}_{right}|m\rangle\otimes|m'\rangle=h_s(m)|m\rangle\otimes|m'\rangle&-\sum_{k=1}^{m}
 \varphi_s({m},{k})
 |m-k\rangle\otimes|m'+k\rangle\,,
 \end{split}
\end{equation}
and
\begin{equation}
\begin{split}
 \mathcal{H}_{left}|m\rangle\otimes|m'\rangle=h_s(m')|m\rangle\otimes|m'\rangle -\sum_{k=1}^{m'}
  \varphi_s(m',{k})
 |m+k\rangle\otimes|m'-k\rangle\,.
 \end{split}
\end{equation}
When acting on the vectors $Y$ the relations above yield
\begin{equation}
\begin{split}
 \langle m,m'|\mathcal{H}_{right}(Y \otimes Y)
 = h_s(m) Y(m)  Y(m')- \sum_{k=1}^{m'} Y(m+k) Y(m'-k)
 \varphi_s({m+k},{k})\,,
 \end{split}
\end{equation}
and
\begin{equation}
\begin{split}
 \langle m,m'|\mathcal{H}_{left}(Y \otimes Y) = h_s(m') Y(m)  Y(m')- \sum_{k=1}^{m} Y(m-k) Y(m'+k)
 \varphi_s({m'+k},{k})\,.
 \end{split}
\end{equation}

On the other hand, inserting $\bar Y$ as given in \eqref{ybar} into the right hand side of \eqref{hyy} we get
\begin{equation}
\begin{split}
 &Y(m)\bar Y(m')-\bar Y(m) Y(m')\\
 &=Y(m) Y(m')\left(h_s(m') -  h_s(m)\right) -Y(m)\sum_{p=1}^{m'} \frac{\oad^{p}}{p}Y(m'-p)+\sum_{p=1}^{m} \frac{\oad^{p}}{p}Y(m-p)Y(m')\,.
 \end{split}
\end{equation}
{Thus \eqref{hyy} is equivalent to
}
\begin{equation}\label{eq:simp1}
\begin{split}
 &-2h_s(m)Y(m) Y(m') -Y(m)\sum_{k=1}^{m'} \frac{\oad^{k}}{k}Y(m'-k)+\sum_{k=1}^{m} \frac{\oad^{k}}{k}Y(m-k)Y(m')\\&=  - \sum_{k=1}^{m'} Y(m+k) Y(m'-k)
 \varphi_s({m+k},{k})- \sum_{k=1}^{m} Y(m-k) Y(m'+k)
 \varphi_s({m'+k},{k})\,.
 \end{split}
\end{equation}
In order to verify the relation above, we note that for our choice of $Y$  we have that
\begin{equation}
\begin{split}
Y(m)  \frac{\oad^{k}}{k}Y(m'-k) =   Y(m+k) Y(m'-k)
 \varphi_s({m+k},{k})\,.
 \end{split}
\end{equation}
This can be seen {when inserting \eqref{eq:YY}} and moving all creation operators to the left. Using the relation above we write \eqref{eq:simp1} as
\begin{equation}
\begin{split}
 &-2h_s(m)Y(m) Y(m')+\sum_{k=1}^{m} \frac{\oad^{k}}{k}Y(m-k)Y(m')=  - \sum_{k=1}^{m} Y(m-k) Y(m'+k)
 \varphi_s({m'+k},{k})\,.
 \end{split}
\end{equation}
Inserting \eqref{eq:YY} and moving the creation operators to the left we find
{
\small
\begin{equation}
\begin{split}
 & \oad^{4s+m+m'} \sum_{k=1}^{m}\kappa(m-k)\kappa(m'+k)\frac{\Gamma\left(\NN+2s+m+m' +1 \right)}{\Gamma\left(\NN +4s+m+m'+1\right)} \frac{\Gamma\left(\NN+m'+k +1 \right)}{\Gamma\left(\NN +2s+m'+k+1\right)}
 \varphi_s({m'+k},{k})\\
 &=
 2h_s(m)\kappa(m)\kappa(m')\oad^{4s+m+m'}\frac{\Gamma\left(\NN +m+m'+2s+1 \right)}{\Gamma\left(\NN+m+m' +4s+1\right)} \frac{\Gamma\left(\NN +m'+1 \right)}{\Gamma\left(\NN +2s+m'+1\right)} \\&\quad-\oad^{4s+m+m'}\sum_{k=1}^{m} \frac{\kappa(m-k)\kappa(m')}{k}\frac{\Gamma\left(\NN +2s+m'+m-k+1 \right)}{\Gamma\left(\NN +4s+1+m'+m-k\right)}\frac{\Gamma\left(\NN +m'+1 \right)}{\Gamma\left(\NN +2s+m'+1\right)}\,.
 \end{split}
\end{equation}}
When multiplying with the appropriate gamma functions from the right, we find that the above equation and thus \eqref{hyy} holds if
\begin{equation}\label{hab}
 2h_s(m)=A+B\,,
\end{equation}
where
\begin{equation}
\begin{split}
 A&= \frac{\Gamma\left(\NN +4s+m+m'+1 \right)}{\Gamma\left(\NN +2s+m+m'+1\right)}\sum_{k=1}^{m} \frac{\kappa(m-k)}{k\kappa(m)}\frac{\Gamma\left(\NN +2s+m-k+m'+1 \right)}{\Gamma\left(\NN +4s+m-k+m'+1\right)}\,,
 \end{split}
\end{equation}
and
\begin{equation}
\begin{split}
 B&=\frac{\Gamma\left(\NN +2s+m'+1 \right)}{\Gamma\left(\NN +m'+1\right)}  \sum_{k=1}^{m}\frac{\kappa(m-k)\kappa(m'+k)}{\kappa(m)\kappa(m')} \frac{\Gamma\left(\NN +m'+k+1 \right)}{\Gamma\left(\NN +2s+m'+k+1\right)}
 \varphi_s({m'+k},{k})\,.
 \end{split}
\end{equation}
The relation \eqref{hab} can be shown as follows. First we write the sums as hypergeometric functions
\begin{equation}
\pFq{4}{3}{a,b,c,d}{e,f,g}{z}=\sum_{n=0}^\infty\frac{(a)_n(b)_n(c)_n(d)_n}{(e)_n(f)_n(g)_n}\frac{z^n}{n!}\,,
\end{equation}
where $(a)_n=\frac{\Gamma(a+n)}{\Gamma(a)}$ denotes the Pochhammer symbol. We obtain
\begin{equation}
\begin{split}
 A
 &=\frac{m  (m+m'+\NN+4 s)}{(m+2 s-1) (m+m'+\NN+2 s)}\textstyle\pFq{4}{3}{1,1,-n_A,a_A}{2,b_A,1+a_A-b_A-n_A}{1}\,,
 \end{split}
\end{equation}
where $n_A=m-1$, $a_A=1-m-m'-\NN-4s$, and $b_A=2-2s-m$ and
\begin{equation}
\begin{split}
 B
 &=\frac{m  (m'+\NN+1) }{(m+2 s-1) (m'+\NN+2 s+1)}\textstyle\pFq{4}{3}{1,1,-n_B,a_B}{2,b_B,1+a_B-b_B-n_B}{1}\,,
 \end{split}
\end{equation}
where $n_B=m-1$, $a_B=2+m'+\NN$, and $b_B=2-2s-m$.
Next, we then use the identity
\begin{equation}\label{hypg}
  \textstyle\pFq{4}{3}{1,1,-n,a}{2,b,1+a-b-n}{1}=\frac{(b-1)(a-b-n)}{(n+1)(a-1)}\left(\psi(n+b)+\psi(1+a-b)-\psi(b-1)-\psi(a-b-n)\right)\,,
\end{equation}
see e.g.~\cite{prudnikov1990integrals}, where $\psi(x)$ denotes the digamma function.
This yields
\begin{equation}
 A=\psi(-m-2 s+1)-\psi(1-2 s)+\psi(-m-m'-\NN-2 s)-\psi(-m'-\NN-2 s)\,,
\end{equation}
and
\begin{equation}
 B=\psi(-m-2 s+1)-\psi(1-2 s)+\psi(m'+\NN+2 s+1)-\psi(m+m'+\NN+2 s+1)\,.
\end{equation}
Finally, using the reflection formula for the digamma function
\begin{equation}\label{refl}
 \psi(1-x)-\psi(x)=\pi \cot (\pi x)\,,
\end{equation}
for   $\NN\in \mathbb{N}$ and  $2s,m,m'\in \mathbb{N}$  we get
\begin{equation}
 A+B=2\left(\psi(m+2s-1)-\psi(2s-1)\right)=2h_s(m)\,,
\end{equation}
which concludes the proof of the bulk algebra.
\paragraph{Boundaries}
The condition for the left boundary \eqref{bndmpa} can be shown to hold in a few lines. We have
\begin{equation}
\begin{split}
 \tilde{\mathcal{B}}_L \lang V|Y&=\sum_{m=0}^\infty \lang V|Y(m)\left(h_s(m)|m\rangle  -\sum_{k=1}^\infty \frac{(\rho_{L}-\rho_R)^k}{k}|m+k\rangle\right)\\
 &= \lang V|\left(\sum_{m=0}^\infty Y(m) h_s(m)|m\rangle  -\sum_{p=1}^\infty\sum_{m=0}^{p-1} Y(m) \frac{(\rho_{L}-\rho_R)^{p-m}}{p-m}|p\rangle\right)\\
 &= \lang V|\sum_{m=1}^\infty\left(  Y(m) h_s(m) -\sum_{p=0}^{m-1} \frac{\oad^{m-p}}{m-p}Y(p)\right)|m\rangle \\
 &= \lang V|\bar Y\,,
 \end{split}
\end{equation}
where in the third step we used that
\begin{equation}
 \lang V|\oad^k =\lang V|(\rho_L-\rho_R)^k\,.
\end{equation}

The relation for the right boundary \eqref{bndmpa} is more tricky. It relies on the identity
\begin{equation}\label{harmid}
\begin{split}
2h_s(m)&=
  \sum_{k=1}^{m} \varphi_s(m,k) \frac{\Gamma(4s+m)\Gamma(2s+m-k) }{\Gamma(4s+m-k)\Gamma(2s+m)}\,.
 \end{split}
\end{equation}
which can be shown to hold when rewriting the right hand side  as
\begin{equation}
\begin{split}
  \sum_{k=1}^{m} \varphi_s(m,k) \frac{\Gamma(4s+m)\Gamma(2s+m-k) }{\Gamma(4s+m-k)\Gamma(2s+m)}=\frac{m (m+4 s-1)}{(m+2 s-1)^2} \textstyle\pFq{4}{3}{1,1,-n_C,a_C}{2,b_C,1+a_C-b_C-n_C}{1}\,,
 \end{split}
\end{equation}
where $n_C=m-1$, $a_C=2-m-4s$ and $b_C=2-2s-m$.
The equality \eqref{harmid} is then shown using \eqref{hypg} and the reflection equation \eqref{refl}. Using \eqref{harmid} it then follows that
\begin{equation}
\begin{split}
   \tilde{\mathcal{B}}_R Y|W\rang &=  \sum_{m=0}^\infty Y(m) h_s(m)|m\rangle|2s-1\rang\\
   &= -\sum_{m=0}^\infty Y(m)\left(  h_s(m)-\sum_{k=1}^{m} \varphi_s(m,k) \frac{\Gamma(4s+m)\Gamma(2s+m-k) }{\Gamma(4s+m-k)\Gamma(2s+m)}|m\rangle\right)|2s-1\rang\\
   &= -\sum_{m=0}^\infty\left( Y(m) h_s(m)-\sum_{k=1}^{m}\frac{\oad^{2s+m}}{k}  \frac{ \Gamma^2(2s+m-k) }{\Gamma(2s)\Gamma(4s+m-k)\Gamma(m-k+1) }|m\rangle\right)|2s-1\rang\\
   &= -\sum_{m=0}^\infty\left( Y(m) h_s(m)-\sum_{k=1}^{m}\frac{\oad^{2s+k} Y(m-k)}{k}  |m\rangle\right)|2s-1\rang\\
   &= -\bar Y|W\rang\,,
   \end{split}
\end{equation}
which concludes this subsection.

\subsection{Similarity transformation and matrix product algebra}\label{sec:sim}

As noted in \eqref{eq:htilde}, a local transformation relates the Hamiltonian \eqref{eq:fullham-b} to the stochastic Hamiltonian \eqref{eq:fullham-sss}. It follows that the matrix product operators $X$ in the steady state \eqref{eq:prodf} are simply related to the operators $Y$ by a rotation
\begin{equation}
 X=\exp\left[-S_-\right] \exp\left[{\rho_R} S_+\right]Y\,,\qquad \bar X=\exp\left[-S_-\right] \exp\left[{\rho_R} S_+\right]\bar Y\,.
\end{equation}
The matrix form of the rotations read
\begin{equation}
\label{eq:rotm}
\langle n'|e^{- S_-} |n\rangle =   (-1)^{n-n'}\binom{n}{n'}\,,\qquad
\langle n'|e^{\rho_R S_+} |n\rangle =  \frac{\rho_R^{n'-n}}{(n'-n)!} \frac{\Gamma(n'+2s)}{\Gamma(n+2s)}\,.
 \end{equation}
Using these formulas we obtain
\begin{equation}
 X(m)=\kappa(m)\oad^{2s}\frac{\Gamma\left(\NN +1 \right)}{\Gamma\left(\NN +2s+1\right)}
 \frac{(\oad+\rho_R)^m}{(1+\oad+\rho_R)^{m+2s}}\,,
\end{equation}
where the fraction of creation operators is interpreted as a power series
\begin{equation}
 \kappa(m)\frac{(\oad+\rho_R)^m}{(1+\oad+\rho_R)^{m+2s}}=\sum_{k=m}^\infty (-1)^{k-m}\binom{k}{m}\kappa(k)(\oad+\rho_R)^k\,.
\end{equation}
The auxiliary operators $\bar X$ can also be obtained but become rather involved.
The boundary vectors remain unchanged.

 We finish this section by presenting  for completeness the matrix product algebra. The bulk relations read
\begin{equation}
\begin{split}
X(m)\bar X(m')-\bar X(m)X(m')&=
 (h_s(m)+h_s(m')) X(m)  X(m')\\
 &\quad\;- \sum_{k=1}^{m'} X(m+k) X(m'-k)
 \varphi_s({m+k},{k})\\
 &\quad\;- \sum_{k=1}^{m} X(m-k) X(m'+k)
 \varphi_s({m'+k},{k})\,,
 \end{split}
\end{equation}
while at the boundaries we have at left
\begin{equation}\label{eq:harmbndss23}
\begin{split}
\lang V|\left[\left(h_s(p)+\sum_{k=1}^\infty\frac{\beta_{L}^k}{k}\right)X(p)-\bar X(p)-\sum_{m=1}^\infty
   \varphi_s({m+p},{m})X(m+p)
-\sum_{k=1}^p X(p-k) \frac{\beta_{L}^k}{k}\right]=0\,,
\end{split}
\end{equation}
and
\begin{equation}\label{eq:harmbndss33}
\begin{split}
\left[\left(h_s(p)+\sum_{k=1}^\infty\frac{\beta_{R}^k}{k}\right)X(p)+\bar X(p)-\sum_{m=1}^\infty
   \varphi_s({m+p},{m})X(m+p)
-\sum_{k=1}^p X(p-k) \frac{\beta_{R}^k}{k}\right]|W\rang=0\,.
\end{split}
\end{equation}
The involved form of the matrix product algebra may explain why a  matrix product solution of the steady state remained unknown up to now.

\section{Conclusion}\label{sec:conc}
In this note we studied the relation between three different types of representations of the steady state of the harmonic process with boundary reservoirs. Besides the closed-form and an integral expression that were known, we derived the matrix product representation which is a new result. Furthermore we derived algebraic relations that constitute the matrix product algebra. Our results relied on the fact that the Markov generator of the harmonic process can be mapped to a triangular (non-stochastic) Hamiltonian by a local similarity transformation. The matrix product solution of the harmonic process is then obtained by applying the similarity transformation to the generators in the matrix product state.

It is well known that the representation of the  matrix product algebra are usually not unique and further equivalent representations can be obtained by performing similarity transformations in the auxiliary space, see e.g.~\cite{wadati} for the case of the ASEP. The same holds true for the harmonic process. Whether a representation exists where the bulk operators in the matrix product state do not depend on the boundary parameters like it is the case in the SEP, is currently not clear. We further remark that the gauge freedom observed in the matrix product representation also exists in the case of the integral representations where the similarity transformations appear as integral transformations. {Throughout the paper we fixed $2s\in\mathbb{N}$ to avoid fractional powers of oscillators, it would be interesting to study carefully if the results can be extended to $2s>0$. Furthermore, it would be interesting to explore the probabilistic structure of matrix product states, cf.~\cite{Temme:2010eif,2025arXiv250916455G}.
}

It would be interesting to derive the matrix product representation presented here directly from the R-matrix following the ideas of \cite{sasamo,Crampee,2015JPhA...48L4001C}, see also \cite{2017JPhA...50d4001K} where a matrix product solution of the q-deformed multispecies version of the process studied here on a closed chain is obtained. Given that for the non-compact representations an integral formulation of the R-matrix exists, see \cite{Derkachov:1999pz}, it seems natural that the steady state, that usually has an infinite-dimensional representation in the auxiliary space, has an integral as well as  a matrix representation. It would be interesting to see if this also holds for processes with finite-dimensional configuration space like the SSEP where so far only the matrix product ansatz and the closed form expression is known.


\subsection*{Acknowledgements}
RF thanks Istv\`{a}n M. Sz\'{e}cs\'{e}nyi for multiple discussions, comments on the manuscript and collaboration on a related topic as well as Cristian Giardin\`a, Eric Ragoucy and Tomohiro Sasamoto for discussions. Further he thanks Davide Gabrielli for pointing out reference \cite{49baa546-5bb8-3d6e-8adc-8b5403de5cf6}, the anonymous referees for reading the manuscript and providing several interesting comments and the group ``\textit{Mathematical Physics of Space, Time and Matter}'' at Humboldt University Berlin for hospitality.
The work was partially supported in part by the INFN grant Gauge and String
Theory (GAST), by the “INdAM–GNFM Project” codice CUP-E53C22001930001, by the FAR
UNIMORE project CUP-E93C23002040005, and by the PRIN project “2022ABPBEY” CUP-E53D23002220006. This research was supported in part by the International Centre for Theoretical Sciences (ICTS) for participating in the program: \textit{Discrete integrable systems: difference equations, cluster algebras and probabilistic models} (code: ICTS/DISDECAP2024/10).

%
%
%
%
%

\bibliography{refs.bib}
\bibliographystyle{utphys2.bst}

\vspace{1cm}
\noindent
%
%
\end{document}